\documentstyle[12pt]{article}
\newcommand{\T}{\textstyle}

\newcounter{tempfigc}			%NEW FIGURE CAPTIONS
\newcommand{\ccaption}[1]{
	\setcounter{tempfigc}{\value{tempfigc}}
	\addtocounter{tempfigc}{1}
        {\centerline{\small {\bf Figure~\thetempfigc:} #1} }}
\newcommand{\fcaption}[1]{
	\setcounter{tempfigc}{\value{tempfigc}}
	\addtocounter{tempfigc}{1}
        {\small
          \begin{description}
          \item[Figure~\thetempfigc:] #1
           \end{description}} }

\begin{document}
%\baselineskip 22 truept

\title{\bf TWO-DIMENSIONAL GRAVITATION AND SINE-GORDON-SOLITONS}

\author{Bernd Stoetzel
\thanks{Supported by Deutsche
Forschungsgemeinschaft. E-mail: Stoetzel@phys.ualberta.ca}\\
Institute for Theoretical Physics\\
Department of Physics, University of Alberta\\
Edmonton, T6G 2J1 Alberta, Canada}

\maketitle

\begin{abstract}
\noindent
Some aspects of two-dimensional gravity coupled to matter fields,
especially to the Sine-Gordon-model are examined. General properties
and boundary conditions of possible soliton-solutions are considered.
Analytic soliton-solutions are discovered and the structure of the
induced space-time geometry is discussed. These solutions have
interesting features and may serve as a starting point for further
investigations.
\end{abstract}

\section{Introduction}

Since the first purely pedagogical study of two-dimensional gravitation
\cite{1} there has been considerable interest in that topic for the
last years [2 - 8]. Apart from being an interesting subject in itself,
two-dimensional models proved to be very useful tools for studying
effects, whose description is rather complicated in realistic
four-dimensional gravitation. Examples are gravitational collapse
\cite{3}, black holes \cite{4}, cosmologies \cite{5} and quantum
effects \cite{6}.

Unfortunately, it is not possible to use the Einstein-equations in two
dimensions, because the Einstein-action is a topological invariant and
therefore, the Einstein-tensor is identically zero. Several suggestions
were made to circumvent this problem: Jackiw \cite{7} and Teitelboim
\cite{8}  proposed a model which they essentially got by dimensional
reduction of the three-dimensional Einstein-action supplemented with a
cosmological constant and an auxiliary field. Mann, et al. \cite{9}
extended this model by giving dynamical content to the auxiliary field.
Another approach was pursued by Callan, et al. \cite{10} who considered
a string-theory inspired action. With the help of a new parameter it is
possible to unite various actions into one,  as was shown by Lemos and
S\'{a} \cite{11}. These different models will be briefly reviewed in
section two.

Much of the analysis of two-dimensional models was concerned with
vacuum-solutions (especially black holes \cite{4}), but it is of course
possible to include also different kinds of matter-couplings to study
the mutual influence of matter and gravity. A very useful example is
the above mentioned model of Callan, et al. \cite{10}, coupled to
massless scalar-fields, because it is exactly soluble on the classical
level. Also, quantum effects, like Hawking-radiation, can be analysed.
It is therefore interesting to see, if any other matter-field models
allow analytic classical solutions as a starting point for
considerations of quantum effects.

As is well-known for many years, there is such a model in flat
space-time, namely the Sine-Gordon-model, whose nonlinear interaction
potential provides classical particle-like kink-solutions \cite{12}. It
proved to be very useful for studying quantization techniques of
nonlinear  baryon-models like the Skyrme-model \cite{13}. Skyrmions
coupled to gravity were analysed recently by Heusler, et al. \cite{14}.
Therefore, it is suggestive to study the  gravitational
Sine-Gordon-model, which is the main purpose of this paper.

In section three some features of the Sine-Gordon-model in flat
space-time and their generalization to curved space-time are discussed.
General aspects of the gravitational Sine-Gordon-model and its
equations of motion are considered in section four and furthermore, the
boundary conditions for possible soliton-solutions are examined. Some
speculations on simple modifications of the Sine-Gordon-model are
presented. In section five various types of analytic solutions are
listed and the structure of the kink and of the space-time for these
different solutions is discussed

\section{Two-dimensional models for gravity}

It is well-known that the Einstein-tensor $G_{\mu\nu}$ vanishes
identically in two dimensions. Therefore, it is not possible to use the
usual Einstein-equations
\begin{equation}
G_{\mu\nu} = 8\pi GT_{\mu\nu}
\end{equation}

\noindent
As a consequence, one has to think about suitable alternatives. Jackiw
\cite{7} and Teitelboim \cite{8} suggested the so-called
constant-curvature-model
\begin{equation}
R+ \Lambda = 0
\end{equation}

\noindent
In four dimensions this is simply the trace of the vacuum
Einstein-equations with cosmological constant. It is useful to have an
action whose variation gives the equations of motion. In four
dimensions such an action is (for the vacuum-equations)
\begin{equation}
I_E = \int \! \! d^4 \! x \sqrt{-g}R
\end{equation}

\noindent
It is also possible to construct a two-dimensional action leading to
(2), but involving an auxiliary scalar-field:
\begin{equation}
I = \int \!  \! d^2 \! x N\sqrt{-g} (R+\Lambda)
\end{equation}

\noindent
Varying $N$ immediately gives (2), whereas the variation with respect
to the metric yields an equation for the field $N$. This action can
also be understood as a dimensionally reduced form of the
three-dimensional Einstein-action. It is then easy to include
matter-interaction in this model by adding a matter-term to the action:
\begin{equation}
I_{JT} = \frac{1}{\kappa} \int \!  \! d^2 \! x N\sqrt{-g} (R+\Lambda +
\kappa {\cal L}_M)
\end{equation}

\noindent
where ${\cal L}_M$ is the matter-lagrangian and $\kappa$ a coupling
constant. In this case the variational equations are
\begin{eqnarray}
N_{;\alpha\beta} - g_{\alpha\beta} {N^{;\mu}}_{;\mu} &=& -{\T
\frac{1}{2}} \kappa N T_{\alpha\beta}\nonumber\\ R+\Lambda &=& -\kappa
{\cal L}_M
\end{eqnarray}

\noindent
with the energy-momentum-tensor defined as
\begin{equation}
T_{\alpha\beta} = g_{\alpha\beta} {\cal L}_M -2 \frac{\delta {\cal
L}_M}{\delta g^{\alpha\beta}}
\end{equation}

\noindent
One important aspect of this model is that the energy-momentum-tensor
is not covariantly conserved.

A natural extension of the Jackiw-Teitelboim-model, which yields a
conserved energy-momentum was suggested by Mann, et al. \cite{9} who
gave dynamical content to the auxiliary field. The action is then
(omitting the cosmological constant)
\begin{equation}
I_M = {1 \over \kappa} \int \! \! d^2 \! x \sqrt{-g}  (-{\T {1 \over
2}} \partial^{\mu} \psi \partial_{\mu} \psi  + \psi R + \kappa {\cal
L}_M )
\end{equation}

\noindent
Instead of (6) the equations of motion now read
\begin{equation}
{\psi^{;\mu}} _{;\mu} = -R
\end{equation}
\begin{equation}
{\psi}_{;\mu} {\psi}_{;\nu} + 2 {\psi}_{;\mu\nu} - {\T {1 \over 2}}
{g}_{\mu\nu} ({\psi}^{;\alpha} {\psi}_{;\alpha} + 4
{\psi^{;\alpha}}_{;\alpha}) = -\kappa {T}_{\mu\nu}
\end{equation}

\noindent
Taking the trace of (10) yields, together with (9),
\begin{equation}
{\psi^{;\mu}}_{;\mu} = {\T {1 \over 2}} \kappa T = -R
\end{equation}

\noindent
This equation is in direct  analogy to the trace of the
four-dimensional Einstein-equation $R \sim T$.

A rather different approach was pursued by Callan, et al. \cite{10} who
used a string-theory inspired action to analyse black hole evaporation
by Hawking radiation:
\begin{equation}
I_{ST} = {1 \over \kappa} \int \! \! d^2 \! x \sqrt{-g} \{
{e}^{-2\varphi} [R + 4{\partial}^{\mu}\varphi\partial_{\mu}{\varphi}] +
\kappa {\cal L}_M \}
\end{equation}

\noindent
Lemos and S\'{a} \cite{11} recently proposed a model which unites all
these different actions into one by introducing a constant parameter:
\begin{equation}
I_{BD} = {1 \over \kappa} \int \! \! d^2 \! x \sqrt{-g}
\{{e}^{-2\varphi} [R - 4\omega{\partial}^{\mu} \varphi {\partial}_{\mu}
\varphi] + \kappa {\cal L}_{M}\}
\end{equation}

\noindent
This is a Brans-Dicke type action in two dimensions \cite{15}. For
$\omega = -1$ one obtains (12) and for $ \omega = 0$ one rediscovers
the Jackiw-Teitelboim model (5) (apart from a slightly different matter
coupling). The model of Mann, et al. is covered by the limit $\omega
\rightarrow \infty$ \cite{16}. The variational equations for this
general action is
\begin{equation}
R = 4\omega ({\varphi^{;\mu}}_{;\mu} - {\varphi}^{;\mu}
{\varphi}_{;\mu})
\end{equation}
\begin{equation}
{\varphi}_{;\mu\nu} - {g}_{\mu\nu} {\varphi^{;\alpha}}_{;\alpha} - 2
(\omega + 1) {\varphi}_{;\mu} {\varphi}_{;\nu} + {g}_{\mu\nu}(\omega+2)
{\varphi}^{;\alpha} {\varphi}_{;\alpha} = {\T {1 \over 4}} \kappa
{e}^{2\varphi} {T}_{\mu\nu}
\end{equation}

\noindent
Which of these different models may be regarded as two-dimensional
general relativity is more or less a matter of taste. In the following
sections I prefer the model of Mann, et al. for several reasons. First,
this model yields as one equation of motion the analogue of the trace
of the Einstein-equations $R \sim T$. Second, it may be obtained as the
limit $\omega \rightarrow \infty$ of the Brans-Dicke action (13) which,
in four dimensions, actually gives the Einstein-model of general
relativity \cite{16}. Third, it can be considered as the $D \rightarrow
2$ limit of $D$-dimensional gravity, as was shown by Mann and Ross
\cite{17}. The fourth reason is rather pragmatic, but  nonetheless
important in the search for analytic solutions: the equations are not
too complicated.

\section{Sine-Gordon-model}

Before considering the coupled system of gravity and Sine-Gordon-model,
first, I would like to review some of the main features of the latter
in flat space-time.

The Sine-Gordon-model is a two-dimensional model for a massless
scalar-field $ \phi $ with a sinus-type self-interaction, given by
\begin{equation}
{\cal L}_{SG} = -{\T {1 \over 2}} {\partial}^{\mu} \phi
{\partial}_{\mu} \phi - U(\phi )
\end{equation}

\noindent
with
\begin{equation}
U(\phi ) = 2{m}^{2} {\sin}^{2} {\T {\phi \over 2}} = {m}^{2} (1 -
\cos{\phi} )
\end{equation}

\noindent
This lagrangian was extensively studied as a tool for testing different
soliton quantization methods, which was of great interest in the study
of solitonic baryon-models, like the Skyrme-model \cite{13}. Another
subject of analysis appeared after the discovery of Coleman \cite{18}
that the Sine-Gordon-model (describing bosons) is equivalent in a
certain sense to the two-dimensional Thirring-model (describing
fermions). Therefore, the Sine-Gordon-model was very fruitful in the
study of different aspects of quantum field theory and there is some
hope that this might also be true for the gravity-coupled
Sine-Gordon-model.

An especially interesting feature of the potential (17) is that it
admits soliton-solutions. Consider the equation of motion for the
lagrangian (16) in the static case
\begin{equation}
{\phi}^{\prime\prime} = {m}^{2} \sin{\phi}
\end{equation}

\noindent
which has the solution
\begin{equation}
\phi = 4\arctan{{e}^{ \pm m (x - {x}_0)}}
\end{equation}

\noindent
This solution connects two successive minima of the potential $U(\phi
)$ or, in other words, two distinct  vacua of the Sine-Gordon-model.
Examining the energy-density for the solution (19) reveals another
important feature of solitons, namely that its energy is bound and
restricted to a certain area of space, which makes it possible to
assign a mass and a size to the soliton. The energy-density is
\begin{equation}
{\cal H} = {\T {1 \over 2}} [{{\phi}^{\prime}}^2 + 2 U \! (\phi )] =
{{\phi}^{\prime}}^2 = {4{m}^{2} \over {\cosh}^{2} [m(x - {x}_{0})]}
\end{equation}

\noindent
$\cal H $ has its maximum at $x = {x}_{0} $, which may therefore be
regarded as the position of the soliton. In the following I take
${x}_{0} = 0 $. The mass of the soliton is
\begin{equation}
M = \int\limits_{-\infty}^{\infty}{\! \!\! dx {\cal H} (x)} = 8m
\end{equation}

\noindent
The most important property of the solution (19) is that one can assign
a conserved soliton-number to it, which may be defined as the spatial
integral over the zeroth component of the obviously conserved current
\begin{equation}
{j}^{\mu} = {1 \over 2\pi} {\epsilon}^{\mu\nu} {\partial}_{\nu} \phi
\end{equation}

\noindent
Where ${\epsilon}^{\mu\nu} $ is the two-dimensional Levi-Civita-symbol
with $ {\epsilon}^{01} = 1 $. For (19), the soliton-number is
\begin{equation}
K = \int\limits_{-\infty}^{\infty}{\!\!\! dx {j}^{0}} = {1 \over 2\pi}
\int\limits_{-\infty}^{\infty}{\!\!\! dx {\phi}^{\prime}} = \pm1
\end{equation}

\noindent
Therefore, the solution with the $+$sign is called soliton and the one
with the $-$sign antisoliton. Concerning the generalization of these
concepts to curved space-time, the soliton-mass is not well defined,
because of the usual problems in defining locally conserved energy in
general relativity, whereas the notion of soliton-number, being a
topological property of the solution, is the same in curved as well as
in flat space-time. The reason for this is that the current (22) stays
to be locally conserved:
\begin{eqnarray}
{{j}^{\mu}}_{;\mu} &=& {1 \over 2\pi} {\epsilon}^{\mu\nu}
{\phi}_{;\nu\mu} = 0\nonumber\\
  &=& {1 \over \sqrt{-g}} (\sqrt{-g} {j}^{\mu} ) _{\prime\mu}
\end{eqnarray}

\noindent
Consequently, one can define the soliton-number in curved space as
\begin{equation}
K = \int\limits_{-\infty}^{\infty}{\!\!\! dx \sqrt{-g} {j}^{0}}
\end{equation}

\section{Gravitational Sine-Gordon-model}

For the reasons discussed in section 2 consider now the following
action describing a gravitational Sine-Gordon-model:
\begin{equation}
I = {1 \over \kappa} \int{\!\!{d}^{2}\! x \sqrt{-g} [ -{\T{1 \over 2}}
{\partial}^{\mu} \psi {\partial}_{\mu} \psi + \psi ( R + \Lambda ) +
\kappa {\cal L}_{SG} (\phi)]}
\end{equation}

\noindent
where ${\cal L}_{SG}$ is the lagrangian (16), $\psi$ is the necessary
auxiliary field as introduced by Mann, et al., $R$ is the usually
defined scalar curvature, g the determinant of the metric tensor
${g}_{\mu\nu}$ ( I use the signature ($-, +, +, +$)), $\Lambda$ is a
cosmological constant  and $\kappa$ is the coupling-constant which is
required to be positive. Otherwise, the action would rather describe an
antigravitational Sine-Gordon-model. An example for a solution with
negative $\kappa$ is given in appendix A. It should be noted that the
matter-gravity coupling in (26) is not conformally invariant. Such type
of a coupling is also considered by Ambj{\o}rn and Ghoroku \cite{19}.
The variation of (26) with respect to $\psi$ and ${g}_{\mu\nu}$ gives
equation (9) and (10) of section 2, which can also be written in the
form (11)
\begin{equation}
R = -{\T{1 \over 2}} \kappa T - {\Lambda}
\end{equation}

\noindent
plus equation (10) for the auxiliary field $\psi$, which is regarded as
a consistency equation without any physical meaning. $T$ in (27) is the
trace of the energy-momentum-tensor defined in (7):
\begin{eqnarray}
{T}_{\mu\nu} & = & {g}_{\mu\nu} {\cal L}_{SG} - 2 {\delta {\cal L}_{SG}
\over \delta {g}^{\mu\nu}}\nonumber\\
& = & {\partial}_{\mu} \phi {\partial}_{\nu} \phi - {\T{1 \over 2}}
{g}_{\mu\nu} [ {\partial}^{\alpha} \phi {\partial}_{\alpha} \phi + 2 U
\! (\phi)]
\end{eqnarray}

\noindent
The trace is therefore
\begin{equation}
T = -2 U \! (\phi)
\end{equation}

\noindent
The scalar-field potential is the source for the metric. One needs of
course also an equation for the scalar-field itself. Variation of the
action with respect to $\phi$ yields
\begin{equation}
{{\phi}^{;\mu}}_{;\mu} = {U}^{\prime} (\phi)
\end{equation}

\noindent
where the prime denotes derivation with respect to $\phi$. With the
help of this equation it is easy to show that the energy-momentum is
covariantly conserved:
\begin{eqnarray}
{{T}^{\mu\nu}}_{;\nu} & = & {({\phi}^{;\mu} {\phi}^{;\nu})}_{;\nu} -
{\T{1 \over 2}} {g}^{\mu\nu} [ {( {\phi}^{;\alpha}
{\phi}_{;\alpha})}_{;\nu} + 2 {U}^{\prime}\! (\phi) {\phi}_{;\nu}]
\nonumber\\
 & = & {\phi}_{;\nu} {\phi}^{;\mu\nu} - {\phi}_{;\nu} {\phi}^{;\nu\mu}
+ {\phi}^{;\mu} {{\phi}^{;\nu}}_{;\nu} - {\phi}^{;\mu} {U}^{\prime}\!
(\phi) \nonumber\\
 & = & 0
\end{eqnarray}

\noindent
Two-dimensional gravity has only one degree of freedom, which means
that the metric is completely characterized by just one function. For
the present purpose it proves to be best to consider a metric of the
following form:
\begin{equation}
{ds}^{2} = -{e}^{2\varphi} {dt}^{2} + {dx}^{2}
\end{equation}

\noindent
Each two-dimensional metric can be brought to that form by a suitable
coordinate transformation. I am interested in static solutions, which
means $\phi = \phi (x)$ and $\varphi = \varphi (x)$. In that case the
scalar curvature for the metric (32) becomes
\begin{equation}
R = -2 ( {\varphi}^{\prime\prime} + {{\varphi}^{\prime}}^{2})
\end{equation}

\noindent
where the prime denotes derivation with respect to $x$. Furthermore,
the second covariant derivative of a scalar-field is
\begin{eqnarray}
{\phi}_{;\mu\nu} & = & {\phi}^{\prime\prime} {\delta}^{x}_{\mu}
{\delta}^{x}_{\nu} - {\varphi}^{\prime} {\phi}^{\prime} {e}^{2 \varphi}
{\delta}^{t}_{\mu} {\delta}^{t}_{\nu} \nonumber\\
{{\phi}^{;\mu}}_{;\mu} & = & {\phi}^{\prime\prime} + {\varphi}^{\prime}
{\phi}^{\prime}
\end{eqnarray}

\noindent
Accordingly, one has to consider the following static equations:
\begin{eqnarray}
{\phi}^{\prime\prime} + \theta {\phi}^{\prime} & = & {U}^{\prime}
(\phi) \\
{\theta}^{\prime} + {\theta}^{2} & = & -{\T{1 \over 2}} \kappa U \! (
\phi ) + {\T{1 \over 2}}\Lambda
\end{eqnarray}

\noindent
with the function $\theta \equiv {\varphi}^{\prime}$. For given
potential $U(\phi)$ these are coupled and nonlinear, but ordinary
differential equations which may, in principle, be solved for given
boundary conditions. A general solution is possible in the easiest case
of a massless scalar-field $U (\phi) = 0$. It is especially interesting
that there exists a soliton-solution in that case, which can be
interpreted as a gravitationally bound kink. This special solution is
discussed in appendix B. Consider now nonlinear potentials like the
Sine-Gordon-potential. What are suitable boundary conditions for a
soliton-solution? As can be seen from the example of the flat-space
Sine-Gordon-model a soliton connects two minima of the potential $U
(\phi)$ and is monotonically rising (19):
\begin{eqnarray}
{\phi}_{-} \leq \phi \leq {\phi}_{+} \, & ; & \hspace{1cm}
{\phi}^{\prime} \geq 0 \nonumber\\
{\phi}^{\prime} ( {\phi}_{\pm} ) = 0 \, & ; & \hspace{1cm} U (
{\phi}_{\pm} ) = {U}_{\pm}
\end{eqnarray}

\noindent
where $ {\phi}_{\pm} = \phi ( x \rightarrow \pm \infty )$ and
${U}^{\prime} ( {\phi}_{\pm} ) = 0$.

To analyse possible solutions of the equations (35) and (36), with the
boundary conditions (37), it is best to decouple the equations. This is
achieved in the following way. Derivation of (36) with respect to $x$
and insertion of (35) yields
\begin{eqnarray}
{\theta}^{\prime\prime} + 2 \theta {\theta}^{\prime} & = & -{\T{1 \over
2}} \kappa {U}^{\prime}\! (\phi) {\phi}^{\prime} = -{\T{1 \over 2}}
\kappa ( {\phi}^{\prime\prime} + \theta {\phi}^{\prime} )
{\phi}^{\prime} \nonumber \\
& = & -{\T{1 \over 4}} \kappa [ {({{\phi}^{\prime}}^{2})}^{\prime} + 2
\theta {{\phi}^{\prime}}^{2} ]
\end{eqnarray}

\noindent
A special solution of this equation obviously is
\begin{equation}
{\theta}^{\prime} = -{\T{1 \over 4}} \kappa {{\phi}^{\prime}}^{2}
\end{equation}

\noindent
If one reinserts this relation into (36) one obtains the function
$\theta$ in dependence of the function $\phi$:
\begin{equation}
{\theta}^{2} = {\T{1 \over 4}} \kappa [ {{\phi}^{\prime}}^{2} - 2 U \!
(\phi) ] + {\T{1 \over 2}}\Lambda
\end{equation}

\noindent
Equation (35), multiplied with ${\phi}^{\prime}$, then becomes an
equation for $\phi$ alone:
\begin{equation}
{1 \over 2} ({{{\phi}^{\prime}}^{2}})^{\prime} \pm {\sqrt{\kappa}
\over 2} {{\phi}^{\prime}}^{2} \sqrt{{{\phi}^{\prime}}^{2} - 2 U \!
(\phi) + {2 \Lambda \over \kappa}} = {[ U (\phi) ]}^{\prime}
\end{equation}

\noindent
where all primes now denote derivation with respect to $x$. This can
also be written as
\begin{equation}
{\sqrt{{{\phi}^{\prime}}^{2} - 2 U \! (\phi) +
\lambda}}^{\hspace{0.1cm}\prime} = \pm {\sqrt{\kappa}  \over 2}
{{\phi}^{\prime}}^{2}
\end{equation}

\noindent
with the rescaled cosmological constant $\lambda$. This seems to be
rather complicated, but may be simplified in the following way. Assume,
there is a solution $\phi = \phi (x)$ and ${\phi}^{\prime} =
{\phi}^{\prime} (x)$. One may now invert the equation for $\phi$ and
insert it in the one for ${\phi}^{\prime}$ to get ${\phi}^{\prime}$ as
a function of $\phi$. Call this function ${F}^{\prime } (\phi)$. Thus,
\begin{equation}
{F}^{\prime} (\phi) \equiv {dF (\phi) \over d \phi} = {\phi}^{\prime}
(x) \equiv {d \phi (x) \over dx}
\end{equation}

\noindent
and
\begin{equation}
{[F (\phi)]}^{\prime} \equiv{d F (\phi) \over dx} = {F}^{\prime}\!
(\phi) {\phi}^{\prime} (x) = {{\phi}^{\prime}}^{2}
\end{equation}

\noindent
Equation (42) can then be written as
\begin{equation}
{\sqrt{{{F}^{\prime}}^{2} - 2 U + \lambda }}^{\hspace{0.1cm}\prime}  =
\pm {\sqrt{\kappa}  \over 2} {[ F(\phi)]}^{\prime}
\end{equation}

\noindent
or, after one integration,
\begin{equation}
\sqrt{{{F}^{\prime}}^{2} - 2U + \lambda} = \pm {\sqrt{\kappa}  \over 2}
F
\end{equation}

\noindent
The integration constant may be set to zero, because $F$ is defined
only up to a constant. Squaring the whole equation then yields a first
order differential equation for the function $F (\phi)$:
\begin{equation}
{{F}^{\prime}}^{2} - {\T {1 \over 4}} \kappa {F}^{2} = 2U - \lambda
\end{equation}

\noindent
It is now possible to translate (37) into conditions for $F$:
\begin{equation}
{F}^{\prime} (\phi) \geq 0 \;  ;  \hspace{1cm} {F}^{\prime}
({\phi}_{\pm})= 0
\end{equation}

\noindent
 From (47) one further may infer
\begin{equation}
{F}^{2} ({\phi}_{\pm}) = -{\T{4 \over \kappa}} ( 2{U}_{\pm} - \lambda)
\end{equation}

\noindent
This means that solutions for positive $\kappa$ are possible if $
2{U}_{\pm} - \lambda \leq 0$. $F (\phi)$ is a monotonically rising
function as well as $\phi (x)$, thus, for the Sine-Gordon-model $(
{U}_{\pm} = 0)$,
\begin{equation}
F ({\phi}_{\pm})  =  \pm {2 \over \sqrt{\kappa}} \sqrt{\lambda}
\end{equation}

\noindent
A soliton-solution of equation (42) is therefore possible for a
positive cosmological constant. What can be said about the metric
function $\theta$? $\theta$ is determined by equation (40) and the
boundary conditions of $\theta$ compatible with that equation are
\begin{equation}
{\theta}^{2} (x  \rightarrow \pm \infty) = - {\T{1 \over 4}} \kappa [
2{U}_{\pm} - \lambda] = {\T{1 \over 4}} \kappa \lambda
\end{equation}

\noindent
where the last equality holds for the Sine-Gordon-model. Solutions with
vanishing cosmological constant are possible if at least one of the two
constants ${U}_{\pm}$ is negative. An example is the following slightly
modified Sine-Gordon-potential:
\begin{equation}
U (\phi) = 2 {m}^{2} {\sin}^{2} {\T {\phi \over 2}} [1 - \alpha \,
{\sin}^{2} {\T{\phi \over 2}}]
\end{equation}

\noindent
Successive minima of this potential are
\begin{equation}
{U}_{-} = 0 \;  ;  \hspace{1cm} {U}_{+} = 2 {m}^{2} ( 1 - \alpha)
\end{equation}

\noindent
and solutions are possible for $\alpha > 1$. In the previous discussion
I used the condition $ {\phi}^{\prime} = 0$, which is true only if $
\phi (x)$ is defined on the whole real line $\Re$. Nevertheless, there
may also be soliton-solutions having a compact domain $I = [{x}_{-},
{x}_{+}]$. This is possible if there are coordinate singularities at
${x}_{\pm}$. In such a case (49) is replaced by
\begin{equation}
{F}^{2} \!  ({\phi}_{\pm}) = -{\T{4 \over \kappa}} (2{U}_{\pm} -
\lambda - {{\T{1 \over 4}} \kappa  {\phi}^{{\prime}^2}_{\pm}})
\end{equation}

\noindent
where $ {\phi}^{\prime}_{\pm} = {\phi}^{\prime} ({\phi}_{\pm})$, and
nothing can be said about the cosmological constant anymore. On the
other hand, (40) shows that ${\phi}^{\prime}_{\pm}$ has to become
infinite, to ensure the coordinate singularities. For that reason there
are no soliton-solutions with compact domain.

\section{Analytic solutions}

Although equation (47) looks rather simple, there is no standard
procedure for solving it and I was not able to determine a general
solution. The only possibility to find at least some special solutions
which may be interpreted as solitons is by trial and error. Consider
first the pure Sine-Gordon-model:
\begin{equation}
{{F}^{\prime}}^{2} - {\T{1 \over 4}} \kappa {F}^{2} = 4{m}^{2}
{\sin}^{2} {\T{\phi \over 2}} - \lambda
\end{equation}

\noindent
The simplest possible ansatz subject to the conditions (48) is
\begin{eqnarray}
{F}^{\prime} \! (\phi) & = & A \sin{\T{\phi \over 2}}\nonumber\\
F (\phi) & = & {F}_{0} - 2 A \cos{\T{\phi \over 2}}
\end{eqnarray}

\noindent
Insertion in (55) yields
\begin{equation}
- {\T{1 \over 4}} \kappa ( {F}^{2}_{0} + 4 {A}^{2}) + \kappa A {F}_{0}
\cos{\T{\phi \over 2}} + {A}^{2} (1 + \kappa) {\sin}^{2} {\T{\phi \over
2}} = -\lambda + 4{m}^{2} {\sin}^{2} {\T{\phi \over 2}}
\end{equation}

\noindent
This equation determines the constants:
\begin{eqnarray}
{F}_{0} = 0 \; ; \hspace{1cm} - \kappa {A}^{2} = -\lambda \nonumber\\
(1 + \kappa) {A}^{2} = 4{m}^{2}
\end{eqnarray}

\noindent
For given Sine-Gordon-parameter $m$ and cosmological constant
$\lambda$, one finds a solution
\begin{equation}
F (\phi) = \pm 2 \sqrt{4{m}^{2} - \lambda} \cos{\T{\phi \over 2} }
\end{equation}
\noindent
but which is a solution only for the special value of the
coupling-constant
\begin{equation}
\kappa = {\lambda \over {4{m}^{2} - \lambda}}
\end{equation}

\noindent
This is consistent with the requirement discussed in the previous
chapter that a positive $\lambda$ is needed to get a solution for
positive $\kappa$. In this special case here one finds an even stronger
requirement, namely
\begin{equation}
0 < \lambda < 4{m}^{2}
\end{equation}

\noindent
The scalar-field is now determined by (43):
\begin{equation}
{\phi}^{\prime} \! (x) = \pm \sqrt{4{m}^{2} - \lambda } \sin{\T{\phi
\over 2}}
\end{equation}

\noindent
which may be integrated at once to give
\begin{equation}
\phi (x) = 4 \arctan {e}^{ \pm M (x - {x}_{0})}
\end{equation}

\noindent
with the soliton mass-parameter
\begin{equation}
M = {\T{1 \over 2}} \sqrt{4{m}^{2} - \lambda }
\end{equation}

\noindent
This has the same form as the flat space kink and antikink solution,
but with a different mass-parameter. One gets the original mass back
for vanishing cosmological constant. As in flat space, ${x}_{0}$ may be
interpreted as the center of the soliton and is set to zero in the rest
of the paper. Also, I concentrate on the kink solution. The discussion
for the antikink is of course completely analogous. Consider now the
metric generated by the soliton. The metric function is given by (39):
\begin{eqnarray}
{\theta}^{\prime} & = &  - {\T{1 \over 4}} \kappa {{\phi}^{\prime}}^{2}
= - \kappa {M}^{2} {\sin}^{2} {\T{\phi \over 2}}\nonumber\\
& = & - {\kappa {M}^{2} \over {\cosh}^{2} (M x)} = -{\lambda \over 4
{\cosh}^{2} (M x)}
\end{eqnarray}

\noindent
and by integration
\begin{equation}
{\varphi}^{\prime} = \theta = -{\lambda \over 4 M} \tanh (M x) +
{\theta}_{0}
\end{equation}

\noindent
The integration-constant ${\theta}_{0}$ is determined by insertion in
the original equations (35, 36), requiring ${\theta}_{0} = 0$. One
further integration then yields
\begin{equation}
\varphi (x) = -{\lambda \over 4 {M}^{2}} \ln \cosh (M x ) = - \kappa
\ln \cosh (M x)
\end{equation}

\noindent
Therefore, the soliton induced metric becomes
\begin{equation}
d{s}^{2} = - {\cosh}^{- 2 \kappa} (M x) d{t}^{2} +d{x}^{2}
\end{equation}

\noindent
This metric is similar to the different metrics analysed by Lemos and
S\'{a} \cite{11} for general two-dimensional dilaton gravity. It is
different, because it does not describe black holes, but the global
structure of the space-time, however, is unexpectedly interesting and
will be discussed below.

There is one point left to complete the examination of  the equations
of motion, namely to check if the consistency equation (10) for the
auxiliary field $\psi$ has a solution. It  is straightforward to show
that this is actually the case with $\psi =  2 \varphi$.

Before analysing the above solution, consider first the modified
Sine-Gordon-potential (52) with vanishing cosmological constant. The
same procedure as in the former case of the pure Sine-Gordon-model
gives now a kink solution of the form
\begin{equation}
\phi (x) = 2 \arctan {e}^{ mx}
\end{equation}
\noindent
and the relation between the model-parameter $\alpha$ and the
coupling-constant $\kappa$ is simply
\begin{equation}
\kappa = 4 ( \alpha - 1)
\end{equation}

\noindent
which again is in accordance with the in section 4 mentioned fact that
$\alpha$ has to be  greater than 1 to make solutions for positive
$\kappa$ possible. The metric for the kink (69) is now
\begin{equation}
d{s}^{2} = - ( 1 + {e}^{2 mx})^{- {\T {\kappa \over 2}}} d{t}^{2} +
d{x}^{2}
\end{equation}

\noindent
I turn now to a detailed discussion of the metric, starting with (68).
The metric is obviously symmetric around the center of the kink and has
coordinate singularities for $x \rightarrow \pm \infty$. From (33) one
finds the scalar curvature
\begin{equation}
R =  {{2 \kappa {m}^{2}} \over {{\cosh}^{2} (M x)}} - \Lambda
\end{equation}

\noindent
Therefore, the curvature is everywhere finite and no physical
singularity arises. Nevertheless, the space-time covered by the
coordinates $t$ and $x$ is not geodesically complete as may be seen
from the geodesic equation
\begin{equation}
{g}_{\alpha\beta} {p}^{\alpha} {p}^{\beta} + {\mu}^{2} = 0
\end{equation}

\noindent
which for the metric (68) reduces to $({p}_{0} = - E)$
\begin{eqnarray}
{\left({dx \over d \lambda}\right)}^{2} & = & {E}^{2} {\cosh}^{2
\kappa} (M x) - {\mu}^{2} \nonumber\\
& \approx & {E}^{2} {( {\T{1 \over 2}} {e}^{\pm Mx})}^{2 \kappa} \; ,
\hspace{1cm} x \rightarrow \pm \infty
\end{eqnarray}

\noindent
The asymptotic behavior is
\begin{equation}
\lambda \sim {e}^{ \mp \kappa Mx} \; , \hspace{1cm} x\rightarrow \pm
\infty
\end{equation}

\noindent
This means that the proper time lapse for a testparticle to travel from
some point ${x}_{0}$ to $x \rightarrow \pm \infty$ is finite. To find
an analytic extension of the solution (68) consider a
Schwarzschild-like coordinate
\begin{equation}
r = \int \! \! dx \, {\cosh}^{- \T \kappa} (M x)
\end{equation}

\noindent
The function $r (x)$ maps the domain $\Re$ of $x$ onto a compact
interval $I$ whose boundary depends on $\kappa$. This provides the
possibility to extend the metric analytically. To get an idea of the
structure of the complete space-time examine two special choices of the
coupling constant $\kappa$, namely $\kappa = 1$ and $\kappa = 2$. These
values are the simplest ones, because the function $r (x)$ becomes
explicitly invertible. For $\kappa = 1$ one obtains
\begin{equation}
r = {\T{2 \over M}} \arctan {e}^{ M x}
\end{equation}

\noindent
The image of this map is the interval $I = [ 0, {\pi \over M}]$ and the
metric has a Schwarzschild-form in the coordinates $r$ and $t$:
\begin{equation}
d{s}^{2} = - {\sin}^{2} (M r) d{t}^{2} + {d{r}^{2} \over {\sin}^{2} (M
r )}
\end{equation}

\noindent
By extending the domain of $r$ from the interval $I$ to $\Re$, one
obtains a geodesically complete manifold. The coordinate-singularities
are now located at the points $Mr = k\pi$ for $k = 0, \pm1, \ldots$. It
is not possible to find a global coordinate system which is
singularity-free (see appendix C). The complete manifold can be
pictured as a "patchwork" of Penrose-diagrams of the form shown in
figure 1.

\begin{figure}[hbt]
\unitlength1cm
\begin{picture}(10,5)(-7,-2.5)

\put(-2,0){\circle*{0.1}} \put(2,0){\circle*{0.1}}
\put(0,2){\circle*{0.1}} \put(0,-2){\circle*{0.1}}
\put(-2,0){\line(1,1){2}} \put(-2,0){\line(1,-1){2}}
\put(2,0){\line(-1,-1){2}} \put(2,0){\line(-1,1){2}}
\put(-2.8,-1.1){$Mr = k \pi$} \put(-2.8,0.9){$Mr = k \pi$}
\put(1.2,0.9){$Mr = (k + 1) \pi$} \put(1.2,-1.1){$Mr = (k+1) \pi$}
\end{picture}

\ccaption{One patch of the Penrose-diagram for the Sine-Gordon-soliton
with $\kappa = 1$}
\end{figure}

In each of these space-time patches exists a soliton or antisoliton
which, in the Schwarzschild-coordinate, has the form
\begin{equation}
\phi (r) = 2 Mr
\end{equation}

\noindent
Note that $Mr$ is restricted for each soliton to one of the intervals
${I}_{k} = [ k\pi, (k+1) \pi]$. By examining the geodesic equation
analogous to (74)
\begin{equation}
{\left({dr \over d\lambda}\right)}^{2} = {E}^{2} - {\mu}^{2} {\sin}^{2}
(Mr)
\end{equation}

\noindent
one finds two types of geodesics. For $E>\mu$ the testparticle will
travel from $r = - \infty$ to $r = \infty$ through infinitely many of
the different patches, whereas for $E < \mu$ it will oscillate between
two neighboring patches, hence, between neighboring solitons.

\noindent
Next consider the case $\kappa = 2$. The Schwarzschild-coordinate is
given by
\begin{equation}
r = {\T{1 \over M}} \tanh (Mx)
\end{equation}

\noindent
The image of the map is now $I = [ - {1 \over M}, {1 \over M}]$ and the
metric has the form:
\begin{equation}
{ds}^{2} = - {(1 - {M}^{2} {r}^{2})}^{2} {dt}^{2} + {{dr}^{2} \over {(1
- {M}^{2} {r}^{2})}^2}
\end{equation}

\noindent
The coordinate-singularities are located at $Mr = \pm 1$. Other
singularities arise at $r \rightarrow \pm \infty$ and these are now
physical singularities as can be seen from the curvature scalar
\begin{equation}
R = 2 \kappa {m}^{2} (1 - {M}^{2} {r}^{2}) - \Lambda
\end{equation}

\noindent
Again it is not possible to find a globally singularity-free coordinate
system. The Penrose-diagram for the extended space-time has the
structure shown in figure 2.

\begin{figure}[hbt]
\unitlength1cm
\begin{picture}(10,10)(-7,-5)
\put(-2,0){\line(1,1){4}} \put(-2,0){\line(1,-1){4}}
\put(-2,4){\line(1,1){0.5}} \put(2,4){\line(-1,1){0.5}}
\put(2,0){\line(-1,-1){4}} \put(2,0){\line(-1,1){4}}
\put(-2,-4){\line(1,-1){0.5}} \put(2,-4){\line(-1,-1){0.5}}
\linethickness{0.1cm}
\put(-2,-4.5){\line(0,1){9}} \put(2,-4.5){\line(0,1){9}}
\put(-3.7,1.9){$r \rightarrow -\infty$} \put(-3.7,-2.1){$r \rightarrow
-\infty$}
\put(2.2,1.9){$r \rightarrow \infty$} \put(2.2,-2.1){$r \rightarrow
\infty$}
\put(-0.9,0.9){$G_-$} \put(-0.9,-1.1){$G_-$} \put(0.4,0.9){$G_+$}
\put(0.3,-1.1){$G_+$}
\end{picture}

\fcaption{Penrose-diagram for the Sine-Gordon-soliton with $\kappa = 1$
($G_\pm$ denote lines with $Mr = \pm 1$ ; bold lines represent physical
singularities)}
\end{figure}

\noindent
The soliton in this case has the form:
\begin{equation}
\phi (r) = 4 \arctan \sqrt{{1 + Mr \over 1 - Mr}}
\end{equation}

\noindent
and is defined only in the regions corresponding to $- 1 \leq Mr \leq
1$. Possible geodesics are oscillations around $Mr = 0$ for $E >\mu$
and around $Mr = \pm 1$ for $E < \mu$. The physical singularities at $
r \rightarrow \pm \infty$ are unreachable for massive particles.
Nevertheless, they are observable, because null geodesics may reach
from $r = - \infty$ to $r = \infty$. In this sense the singularities in
this space-time are naked. There is no horizon hiding them.

In the remaining part of this section I will discuss the metric (71),
induced by the soliton (69) of the modified Sine-Gordon-potential (52)
with vanishing cosmological constant.

\noindent
Calculating the kink-number (25) yields in this case
\begin{equation}
K = {1 \over 2\pi} \int\limits_{- \infty}^{\infty} \!\!\! dx
{\phi}^{\prime} = {1 \over 2\pi} [ \phi (\infty) - \phi ( - \infty)] =
{1 \over 2}
\end{equation}

\noindent
and one might expect an extension of the space-time which completes
also the kink. The same conclusion may be drawn from the fact that the
space-time covered by $t$ and $x$ is geodesically incomplete for $ x
\rightarrow \infty$, because the proper time lapse for testparticles
approaching this limit is finite for the same reason as in the previous
case. Also, in this limit the metric (71) is singular, but the
space-time is not. The scalar curvature is finite at this end of the
real line:
\begin{equation}
R = {\kappa {m}^{2} \over 2 {\cosh}^{2} (Mx)} (1 - {\T{\kappa \over 4}}
{e}^{ 2 mx}) = - {{\kappa}^{2} {m}^{2} \over 2} \; ; \hspace{1cm} x
\rightarrow \infty
\end{equation}

\noindent
On the other side of the real line, $x \rightarrow -\infty$, the
curvature vanishes, $R = 0$, and the space-time is asymptotically flat.

\noindent
To find an analytic extension switch again to a
Schwarzschild-coordinate defined by
\begin{equation}
r = \int \!\! dx (1 + {e}^{2 mx})^{-{\kappa \over 4}}
\end{equation}

\noindent
The image of the function $r (x)$ is now the half-open interval $I = (
- \infty, 0]$ and an extension into the region $r > 0$ is possible.
Again the specific structure of the complete space-time depends on the
value of $\kappa$ and as examples I consider in the following the cases
$\kappa = 2$ and $\kappa = 4$.

\noindent
For $\kappa = 2$ the coordinate transformation (87) is explicitly
\begin{equation}
r (x) = {1 \over 2 m} \ln {\sqrt{1 + {e}^{2 mx}} - 1  \over \sqrt{1 +
{e}^{2 mx}} + 1}
\end{equation}

\noindent
which yields the metric
\begin{equation}
{ds}^{2} = - {\tanh}^{2} (mr) {dt}^{2} + {\coth}^{2} (mr) {dr}^{2}
\end{equation}

\noindent
Now the space-time is asymptotically flat at both sides of the real
line and the structure of the space-time is analogous to an extreme
Kerr-geometry given by the Penrose-diagram of figure 3 \cite{20}.

\begin{figure}[htb]
\unitlength1cm
\begin{picture}(10,10)(-6.5,-3.5)
\thicklines
\put(-2,0){\circle*{0.1}}  \put(2,0){\circle*{0.1}}
\put(0,2){\circle*{0.1}} \put(-2,4){\circle*{0.1}}
\put(2,4){\circle*{0.1}} \put(4,2){\circle*{0.1}}
\put(0,-2){\circle*{0.1}} \put(4,-2){\circle*{0.1}}
\put(-2,0){\line(1,1){2}} \put(-2,0){\line(1,-1){2}}
\put(2,0){\line(1,1){2}} \put(2,0){\line(1,-1){2}}
\put(0,2){\line(-1,1){2.5}} \put(0,-2){\line(-1,-1){0.5}}
\put(4,2){\line(-1,1){2.5}} \put(4,-2){\line(-1,-1){0.5}}
\thinlines
\put(0,2){\line(1,1){2.5}} \put(0,2){\line(1,-1){2}}
\put(2,0){\line(-1,-1){2}} \put(0,-2){\line(1,-1){0.5}}
\put(-2,4){\line(1,1){0.5}}
\put(-2.8,0.9){$r \rightarrow - \infty$} \put(-2.8,-1.1){$r \rightarrow
- \infty$}
\put(3.3,0.9){$r \rightarrow + \infty$} \put(3.3,2.9){$r \rightarrow +
\infty$}
\put(1.1,1.1){$G_0$} \put(1.1,-1.2){$G_0$}
\end{picture}

\fcaption{Penrose-diagram for the modified Sine-Gordon-soliton with
$\kappa = 2$ ($G_0$  denotes lines with $mr = 0$)}
\end{figure}

\noindent
The soliton in these coordinates has the form
\begin{equation}
\phi (r) = 2 \arctan {\sinh}^{- 1} (mr)
\end{equation}

\noindent
and the kink-number is $K = 1$ as expected. Geodesics range from $r = -
\infty$ to $r = + \infty$ for $E > \mu$ and oscillate around $r = 0$
for $E < \mu$.

\noindent
For $\kappa = 4$ one obtains
\begin{equation}
r (x) = - {1 \over 2 m} \ln (1 + {e}^{- 2 mr})
\end{equation}

\noindent
and the metric in this case is
\begin{equation}
{ds}^{2} = - {(1 - {e}^{2 mr})}^{2} {dt}^{2} + {{dr}^{2} \over {(1 -
{e}^{2 mr})}^{2}}
\end{equation}

\noindent
Apart from $r = 0 (x \rightarrow \infty)$ another singularity arises at
$r \rightarrow \infty$ which, after calculating the curvature
\begin{equation}
R = 8 {m}^{2} {e}^{2 mr} (1 - 2 {e}^{2 mr})
\end{equation}

\noindent
appears to be a true physical singularity. The space-time is now
analogous to an extreme Reissner-Nordstr\"{o}m-geometry \cite{21}. The
Penrose-diagram is shown in figure \nolinebreak 4.

\begin{figure}[hbt]
\unitlength1cm
\begin{picture}(10,10)(-7.5,-3)

\put(-2,0){\circle*{0.1}} \put(2,0){\circle*{0.1}}
\put(-2,4){\circle*{0.1}} \put(2,4){\circle*{0.1}}
\put(0,2){\circle*{0.1}} \put(0,-2){\circle*{0.1}}
\thicklines
\put(-2,0){\line(1,1){2}} \put(-2,0){\line(1,-1){2}}
\put(-2,4){\line(1,-1){2}}
\put(-2,4){\line(1,1){0.5}} \put(0,-2){\line(-1,-1){0.5}}
\thinlines
\put(0,2){\line(1,1){2}} \put(0,2){\line(1,-1){2}}
\put(2,0){\line(-1,1){2}} \put(2,0){\line(-1,-1){2}}
\put(2,4){\line(-1,1){0.5}} \put(0,-2){\line(1,-1){0.5}}
\linethickness{0.1cm}
\put(2,-2.5){\line(0,1){7}}
\put(-2.8,0.9){$r \rightarrow  -\infty$} \put(-2.8,-1.1){$r \rightarrow
-\infty$}
\put(2.2,1.9){$r \rightarrow  \infty$} \put(0.3,-1.1){$G_0$}
\put(0.3,0.9){$G_0$}
\end{picture}

\fcaption{Penrose-diagram for the modified Sine-Gordon-soliton with
$\kappa = 4$ ($G_0$  denotes lines with $mr = 0$; bold line represent
the physical singularity)}
\end{figure}

\noindent
The soliton is given by
\begin{equation}
\phi (r) = 2 {\cot}^{- 1} \sqrt{{e}^{- 2 mr} - 1}
\end{equation}

\noindent
which is not defined in the region $r > 0$. Therefore the soliton is
not completed in this case. The missing half of it is cut off by the
space-time singularity at $r \rightarrow \infty$. Geodesics range from
$r = - \infty$ to ${r}_{max} = {1 \over 2m} \ln (1 + {E \over \mu})$
for $E \geq \mu$ and oscillate around $r = 0$ for $E< \mu$. The
space-time singularity is unreachable for massive testparticles, but
observable by light-rays.

\section{Conclusion}

The purpose of this work was to examine some aspects of two-dimensional
gravity coupled to matter-fields with a special emphasize on the
Sine-Gordon-model. There is no unique model for two-dimensional gravity
and I used the one proposed by Mann, et al., because it seems to be the
most natural analogue to general relativity in four dimensions. General
properties of solitonic solutions in this model were discussed and
different analytic solutions were found. The examination of the
appropriate geometries showed very interesting and unexpected
space-time-structures for these solutions.

Although I was not able to find other analytic solutions than the ones
discussed in the above sections, this, of course, does not mean that
there are no other solutions. It would be especially interesting to
find soliton-solutions with a black hole metric, because a great part
of the work done in two-dimensional gravity is connected with questions
of black hole physics, like for example Hawking-radiation.

The Sine-Gordon-model was the subject of many investigations concerning
quantum field theory in flat space-time and I hope that the solutions
discussed in the present work may be useful starting points for similar
investigations in quantum gravity.

After finishing my work I became aware of a recent preprint of Shin and
Soh \cite{22} dealing also with Sine-Gordon-solitons in two-dimensional
gravity. They took the gravity-model of Callan, et al., and were able
to discover a soliton-solution with a black hole metric, because the
matter-gravity coupling they used is slightly different from the one I
used in my work.

\section*{Acknowledgments}
I wish to thank my wife Simone Stoetzel for type-setting the
manuscript.

\pagebreak

{\LARGE \centerline{\bf Appendices}}
\vspace{2cm}

\begin{appendix}
\section{Antigravitational soliton}

It was mentioned in section 4 that the action (26) with negative values
of $\kappa$ gives rise to an antigravitational model. In this appendix
I give an example of such a soliton-solution for the pure
Sine-Gordon-potential with vanishing cosmological constant. The same
methods used in section 6 reveal the following solution for $\kappa = -
4$ :
\begin{equation}
\phi (x) = 2 \arctan {e}^{ mx}
\end{equation}

\noindent
Actually, this is just one half of the kink, because it does not
connect two successive minima (vacua) of the potential, but a minimum
and an extremum. As in the last example of section 5 one might again
expect an extension of space-time to complete the kink or a physical
singularity to cut off the second half of the soliton.

\noindent
The induced metric is
\begin{equation}
{ds}^{2} = - {(1 + {e}^{2 mx})}^{2} {dt}^{2} + {dx}^{2}
\end{equation}

\noindent
Note that this is in accordance with the metric solution (71) for the
modified Sine-Gordon-potential for $ \alpha = 0 $, as it has to be.
Introduce now a null-coordinate
\begin{equation}
{x}^{*} = - {1 \over 2 m} \ln (1 + {e}^{- 2 mx})
\end{equation}

\noindent
which gives the conformally flat metric
\begin{equation}
{ds}^{2} = - {(1 - {e}^{{2 mx}^{*}})}^{- 2} ({dt}^{2} - {dx}^{* 2})
\end{equation}

\noindent
The original domain of $x$ is mapped on the half-line $( - \infty, 0]$
and the space-time may be extended into the region ${x}^{*} > 0$. The
curvature is
\begin{equation}
R = - 8 {m}^{2} {e}^{{2 m x}^{*}}
\end{equation}

\noindent
thus revealing a physical singularity at ${x}^{*} \rightarrow \infty$
and a coordinate singularity at $x^* = 0$ (resp. $x \rightarrow
\infty$). The Penrose-diagram  for this case is shown in figure 5.

\begin{figure}[hbt]
\unitlength1cm
\begin{picture}(10,6)(-7.5,-2.5)

\put(-2,0){\circle*{0.1}} \put(2,0){\circle*{0.1}}
\put(0,2){\circle*{0.1}} \put(0,-2){\circle*{0.1}}
\thicklines
\put(2,0){\line(-1,1){2}} \put(2,0){\line(-1,-1){2}}
\put(2.05,0){\line(-1,1){2}} \put(2.05,0){\line(-1,-1){2}}
\thinlines
\put(-2,0){\line(1,1){2}} \put(-2,0){\line(1,-1){2}}
\multiput(0,-2)(0,0.4){10}{\line(0,1){0.3}}
\put(-2.9,0.9){$x^* \rightarrow  -\infty$} \put(-2.9,-1.1){$x^*
\rightarrow  -\infty$}
\put(1.2,0.9){$x^* \rightarrow  \infty$} \put(1.2,-1.1){$x^*
\rightarrow  \infty$}
\put(0.2,-0.1){$G_0$}
\end{picture}

\fcaption{Penrose-diagram for antigravitational soliton in the
Sine-Gordon-model with vanishing cosmological constant. ($G_0$ denotes
line with $x^* = 0$; bold lines represent the physical singularity)}
\end{figure}

\noindent
The soliton in the new coordinate is
\begin{equation}
\phi ({x}^{*}) = 2 {\cot}^{-1} \sqrt{{e}^{-2 m {x}^{*}} - 1}
\end{equation}

\noindent
which is not defined in the region ${x}^{*} > 0$. Again the missing
half of the kink is cut off by the singularity.

\noindent
The geodesic equation is
\begin{equation}
\left({{{dx}^{*} \over d \lambda}}\right)^{2} = [ {E}^{2} {(1 - {e}^{2
m {x}^{*}})}^{2} - {\mu}^{2}] {(1 - {e}^{2 m{x}^{*}})}^{2}
\end{equation}

\noindent
Massive testparticles stay always in one of the two halves of
space-time depending on their initial conditions. Especially the point
${x}^{*} = 0$ (which may be interpreted as the center of the soliton)
can never be reached for $\mu \neq 0$. This shows the antigravitational
character of the solution. The soliton repels every massive object.

\section{Gravitationally bound soliton}

In this appendix I discuss a solution of the equations (35, 36) which
can be interpreted as a gravitationally bound soliton. For a massless
scalar-field the potential vanishes and equation (36) is easily solved
by
\begin{equation}
\theta (x) = \lambda \tanh (\lambda x)
\end{equation}

\noindent
with the parameter $\lambda = \sqrt{{\Lambda \over 2}}$. This is a real
solution for positive cosmological constant. Equation (35) yields
\begin{equation}
\phi (x) = {\phi}_{0} \arctan {e}^{ \lambda x}
\end{equation}

\noindent
with an arbitrary constant ${\phi}_{0}$. The appropriate metric is
given by
\begin{equation}
{ds}^{2} = - {\cosh}^{2} (\lambda x) d{t}^{2} + d{x}^{2}
\end{equation}

\noindent
This is the anti-de Sitter geometry \cite{23}. The space-time consists
of infinitely many patches all filled by one soliton of the form (103)
and completely separated for massive objects, which oscillate around
one kink. Nevertheless, the different patches are observable with
light-rays, because null-geodesics may reach from one to another part
of the space-time.

\section{Treatment of coordinate singularities}

The usual treatment of coordinate singularities is patterned after the
method of Kruskal introduced in dealing with the Schwarzschild-metric
in four dimensions \cite{24}. In two dimensions the recipe is the
following: starting with the Schwarzschild-type metric, like for
example (78), transform to a conformally flat coordinate system in such
a way that the singularities disappear.

\noindent
A generic Schwarzschild-type metric is
\begin{equation}
{ds}^{2} = - {F}^{2} (r) {dt}^{2} + {{dr}^{2} \over {F}^{2} (r)}
\end{equation}

\noindent
A transformation to a conformally flat system is obviously achieved by
\begin{equation}
{r}^{*} = \int \! \! {dr \over F (r)}
\end{equation}

\noindent
The resulting metric is
\begin{equation}
{ds}^{2} = -{{F}^{*}}^{2} \! ({r}^{*}) ({dt}^{2} - {{dr}^{*}}^{2})
\end{equation}

\noindent
The singularity is at some point ${r}^{*} = {r}^{*}_{0}$. Find now a
transformation to get rid of  it:
\begin{equation}
U = G (t, {r}^{*}) \; , \hspace{1cm} V = H (t, {r}^{*})
\end{equation}

\noindent
The condition that the metric stays conformally flat yields conditions
for the functions $H$ and $G$:
\begin{equation}
\dot{H} = {G}^{\prime} \; ; \hspace{1cm} {H}^{\prime} = \dot{G}
\end{equation}

\noindent
with the general solutions
\begin{eqnarray}
G (t, {r}^{*}) & = & g ({r}^{*} + t) + h ({r}^{*} - t)\nonumber\\
H (t, {r}^{*}) & = & g ({r}^{*} + t) - h ({r}^{*} - t)
\end{eqnarray}

\noindent
and the new metric is
\begin{equation}
{ds}^{2} = {{{F}^{*}}^{2} \! ({r}^{*}) \over 4 {g}^{\prime} ({r}^{*} +
t) {h}^{\prime} ({r}^{*} - t)} ({dU}^{2} - {dV}^{2})
\end{equation}

\noindent
Therefore, the function $\Omega ({r}^{*}, t) = {g}^{\prime} ({r}^{*} +
t) {h}^{\prime} ({r}^{*} - t)$ has to cancel the singularity in
${F}^{*} ({r}^{*})$, which means
\begin{equation}
 \Omega ({r}^{*}, t) \stackrel{{r}^{*} \rightarrow {r}^{*}_{0}}
\longrightarrow {{F}^{*}}^{2} ({r}^{*}_{0}) \; , \hspace{1cm} t =
const.
\end{equation}

\noindent
For finite ${r}^{*}_{0}$ this is only possible if $\dot{\Omega} = 0$
which is equivalent to $\Omega ({r}^{*}) \sim {e}^{\gamma {r}^{*}}$. An
example for this is the antigravitational solution of appendix A, where
${r}^{*}_{0} = 0$ (${r}^{*}$ is called ${x}^{*}$ in that section), but
near that value the conformal factor is
\begin{equation}
{{F}^{*}}^{2} ({r}^{*}) \stackrel{{r}^{*} \rightarrow 0}
\longrightarrow {1 \over {(2 m {r}^{*})}^{2}}
\end{equation}

\noindent
Therefore, this singularity cannot be removed by any coordinate
transformation of the required type.

In all other examples of this paper ${r}^{*}_{0}$ is infinite. In that
case, one has not to impose the additional condition $\dot{\Omega} = 0$
and there is some more freedom than in the previous case. Nevertheless,
the behavior of all these metrics near the singularity is of the form
\begin{equation}
{ds}^{2} \sim {- {dt}^{2} + d{{r}^{*}}^{2} \over 4 {m}^{2}
{{r}^{*}}^{2}} \; ; \hspace{1cm} {r}^{*} \rightarrow \infty
\end{equation}

\noindent
Thus, one needs functions $g$ and $h$ such that
\begin{equation}
{g}^{\prime} ({r}^{*} + t) {h}^{\prime} ({r}^{*} - t) \sim {1 \over
{{r}^{*}}^{2}}
\end{equation}

\noindent
for ${r}^{*} \rightarrow \infty$ and $ t = const.$ or, in the same
limit,
\begin{eqnarray}
{g}^{\prime} ({r}^{*} + t) & \sim & {1 \over {({r}^{*} +
t)}^{\alpha}}\nonumber\\
{h}^{\prime} ({r}^{*} - t) & \sim & {1 \over {({r}^{*} - t)}^{2 -
\alpha}}
\end{eqnarray}

\noindent
which means for the functions themselves $(\alpha \neq 1)$:
\begin{eqnarray}
g ({r}^{*} + t) & \sim & {1 \over {({r}^{*} + t)}^{\alpha -
1}}\nonumber\\
h ({r}^{*} - t) & \sim & {1 \over {({r}^{*} - t)}^{1 - \alpha}}
\end{eqnarray}

\noindent
In terms of the new coordinates
\begin{eqnarray}
U & \sim & {({r}^{*})}^{1 - \alpha} + {({r}^{*})}^{\alpha -
1}\nonumber\\
V & \sim & {({r}^{*})}^{1 - \alpha} - {({r}^{*})}^{\alpha - 1}
\end{eqnarray}

\noindent
If ${r}^{*}$ approaches the singularity both, $U$ and $V$, diverge;
hence, the new coordinate-system does not cover the whole space-time
described by the coordinates ${r}^{*}$ and $t$. The last possibility
left is the choice $\alpha = 1$ in (116). In that case, one obtains
\begin{equation}
U \sim \ln ({r}^{*}) \; ; \hspace{1cm} V\sim \ln ({r}^{*})
\end{equation}

\noindent
and again $U$ and $V$ diverge at the singularity. In other words, it is
not possible to find a globally singularity-free system, if the
singularity is of the form (114).

\end{appendix}


\begin{thebibliography}{99}
\bibitem{1} P. Callas; {\it Am. J. Phys.} {\bf 45} (1977) 833
\bibitem{2} J. D. Brown; {\it Lower Dimensional Gravity} (Singapore:
World Scientific 1988)
\bibitem{3} R. B. Mann, S. F. Ross; {\it Class. Quant. Grav.} {\bf 9}
(1992) 2335\\
                    R. B. Mann; {\it Report} No. WATPHYS TH-91/06 (unpublished)
\bibitem{4} E. Witten; {\it Phys. Rev. D} {\bf 44} (1991) 314\\
                    C. G. Callan, S. B. Giddings, J. A. Harvey, A. Strominger;
{\it Phys.Rev. D} {\bf 45} (1992) R1005\\
                    R. B. Mann, A. Shiekh, L. Tarasov; {\it Nucl. Phys. B}
{\bf 341} (1990)134\\
                    R. B. Mann, M. S. Morris, S. F. Ross; {\it Class. Quant.
Grav} {\bf 10}(1993) 1477\\
                    R. B. Mann, S. M. Morsink, A. E. Sikkema, T. G. Steele;
{\it Phys. Rev.D} {\bf 43} (1991) 3948\\
                    J. P. S. Lemos, P. M. S\'{a}; {\it Mod. Phys. Lett. A}
{\bf 9} (1994)771\\
                    J. P. S. Lemos, P. M. S\'{a}; {\it Phys. Rev. D}{\bf 49}
(1994) 2897\\
                    J. P. S. Lemos; {\it Class.Quant.Grav.}{\bf 12}
(1995) 1081 \\
                    R. Balbinot, P. R. Brady; {\it Class. Quant. Grav}
{\bf 11} (1994) 1763\\
                    J. Y. Kim, H. W. Lee, Y. S. Myung; {\it Phys. Lett.
B} {\bf 328} (1994) 291\\
                    A. Kumar, K. Ray; {\it Preprint} IP/BBSR/94-54,
hep-th/9410068 (1994)\\
                    R. C. Myers; {\it Phys.Rev. D}{\bf 50} (1994) 6412
\bibitem{5} J. S. F. Chan, R. B. Mann; {\it Class. Quant. Grav.} {\bf
12} (1995) 351\\
                    A. E. Sikkema, R. B. Mann; {\it Class. Quant.
Grav.} {\bf 8} (1991) 219
\bibitem{6} J. Ambj{\o}rn, K. Ghoroku; {\it Int. J. Mod. Phys. A} {\bf
9} (1994) 5689\\
                    N. Sanchez; {\it Nucl. Phys. B} {\bf 266} (1986)
487\\
                    R. Floreanini; {\it Ann. Phys.} {\bf 167} (1986)
317\\
                    K. G. Akdeniz, \"{O}. F. Dayi, A. Kizilers\"{u};
{\it Mod. Phys. Lett. A} {\bf 7} (1992) 1757\\
                    S. W. Hawking, J. D. Hayward; {\it Phys. Rev. D}
{\bf 49} (1994) 5252\\
                    Y. Kazama, Y. Satoh, A. Tsuchiya; {\it Phys. Rev. D}
{\bf 51} (1995) 4265
\bibitem{7} R. Jackiw; in {\it Quantum Theory of Gravity}, ed. S.
Christensen (Bristol: Adam Hilger 1984) p. 403\\
                    R. Jackiw; {\it Nucl. Phys. B} {\bf 252} (1985) 343
\bibitem{8} C. Teitelboim; in {\it Quantum Theory of Gravity}, ed. S.
Christensen (Bristol: Adam Hilger 1984) p. 327
\bibitem{9} R. B. Mann, S. M. Morsink, A. E. Sikkema, T. G. Steele;
(see ref. 4)
\bibitem{10} C. G. Callan, S. B. Giddings, J. A. Harvey, A. Strominger;
(see ref. 4)
\bibitem{11} J. P. S. Lemos, P. M. S\'{a}; {\it Phys. Rev.} (see ref. 4)
\bibitem{12} R. Rajaraman; {\it Solitons and Instantons} (Amsterdam:
North Holland 1982)
\bibitem{13} R. Jackiw; {\it Rev. Mod. Phys.} {\bf 49} (1977) 681\\
                      C. Rebbi, G. Soliani; {\it Solitons and
Particles} (Singapore: World Scientific 1984)
\bibitem{14} M. Heusler, N. Straumann, Z. Zhou; {\it Helv. Phys. Acta}
{\bf 66} (1993) 614\\
                       M. Heusler, S. Droz, N. Straumann; {\it Phys.
Lett. B} {\bf 285} (1992) 21\\
                       M. Heusler, S. Droz, N. Straumann; {\it Phys.
Lett. B} {\bf 271} (1991) 61\\
                       S. Droz, M. Heusler, N. Straumann; {\it Phys.
Lett. B} {\bf 268} (1991) 371
\bibitem{15} Ch. W. Misner, K. S. Thorne, J. A. Wheeler; {\it
Gravitation} (San Francisco: W. H. Freeman and Company 1973)
\bibitem{16} J. P. Lemos, P. M. S\'{a}; {\it Class. Quant. Grav.} {\bf
11} (1994) L11
\bibitem{17} R. B. Mann, S. F. Ross; {\it Class. Quant. Grav.} {\bf 10}
(1993) 1405
\bibitem{18} S. Coleman; {\it Phys. Rev. D} {\bf 11} (1975) 2088
\bibitem{19} J. Ambj{\o}rn, K. Ghoroku; (see ref. 6)
\bibitem{20} B. Carter; {\it Phys. Rev.} {\bf 141} (1966) 1242
\bibitem{21} B. Carter; {\it Phys. Lett.} {\bf 21} (1966) 423
\bibitem{22} H.-S. Shin, K.-S. Soh; {\it Preprint} hep-th/9501045
(1995) (to appear in {\it Phys. Rev. D})
\bibitem{23} M. Ba\~{n}ados, M. Henneaux, C. Teitelboim, J.
Zanelli; {\it Phys. Rev. D} {\bf 48} (1993) 1506
\bibitem{24} J. C.
Graves, D. R. Brill; {\it Phys. Rev.} {\bf 120} (1960) 1506

\end{thebibliography}
\end{document}